\documentclass[manuscript]{aastex}

\slugcomment{Not to appear in Nonlearned J., 45.}

\shorttitle{Study of neutron-capture element abundances}
\shortauthors{Li et al.}

\begin{document}

\title{Study of Neutron-Capture Element Abundances in Metal-Poor Stars}


\author{Hongjie Li\altaffilmark{1,2}, Xiaojing Shen\altaffilmark{1}, Shuai Liang\altaffilmark{1}, Wenyuan Cui\altaffilmark{1} and Bo Zhang\altaffilmark{1,3}}

\affil{1. Department of Physics and Hebei Advanced Thin Films Laboratory, Hebei Normal University, Shijiazhuang 050016, China\\
2. School of Sciences, Hebei University of Science and Technology,
Shijiazhuang 050018, China}

\altaffiltext{3}{Corresponding author. E-mail
address: zhangbo@mail.hebtu.edu.cn}

\begin{abstract}

This work describes a study of elemental abundances for 30
metal-poor stars whose chemical abundances provide excellent
information for setting constraints on models of neutron-capture
processes. Based on the abundances of main r- and weak r-process
stars, the abundance patterns of main r-process and weak r-process
are obtained. The two r-process component coefficients are defined
to determine the relative contributions from individual
neutron-capture process to abundances of metal-poor stars. Based on
the component coefficients, we find that metal-poor stars BD+4 2621
and HD 4306 are also weak r-process stars, which means that the
abundance pattern produced by weak r-process is stable. All
metal-poor star abundances contain the contributions of both main
r-process and weak r-process. The elements produced by weak
r-process have increased along with Fe over the polluted history.
Most of the metal-poor star abundances do not follow the pattern
observed in solar system, but there is a small fraction that do. For
the low-[Sr/Fe] star BD-18 5550 ([Sr/Fe]$\lesssim-1$),
neutron-capture element abundances can be explained by the mixture
of two r-process components. Since lighter elements in this star
cannot be fitted by the two components, the abundance pattern of
P-component is estimated from those abundances.

\end{abstract}


\section{Introduction}

There are two distinct neutron-capture processes responsible for the
generation of the elements heavier than the Fe peak elements: the
slow neutron-capture process (the s-process) and the rapid
neutron-capture process (the r-process) \citep{bur57}. The s-process
is further divided into two categories: the weak s-component and the
main s-component. Massive stars are sites of the weak component of
s-process nucleosynthesis, which is mainly responsible for the
production of lighter neutron-capture elements (e.g. Sr, Y, and Zr)
\citep{lam77,rai91,rai93,the00}. The s-process contribution to
heavier neutron-capture elements (heavier than Ba) is due only to
the main s-component. The low- to intermediate-mass
($\approx$1.3$-$8$M_\odot$) stars in the asymptotic giant branch
(AGB) are usually considered to be sites in which the main s-process
occur \citep{bus99}. \citet{arl99} derived the ``residual r-process
abundances" for the solar system by subtracting the s-process
abundances from the total abundances in the solar system. There is
abundant evidence suggesting that Type II supernovae (SNe II) are
sites for the synthesis of the r-process nuclei, although this has
not yet been fully confirmed \citep{cow91,sne08}. The observations
and analysis on very metal-poor stars imply that the stars with
[Fe/H]$\leq-2.5$ might form from gas clouds polluted by a few
supernovae (SNe) \citep{mcw95a,mcw95b,rya96}. Therefore, the
abundances of heavy elements in metal-poor stars have been used to
learn about the nature of the nucleosynthetic processes in the early
Galaxy.

Because of the large overabundance of heavier neutron-capture
elements (Z$\geq56$) relative to iron ([Eu/Fe]$\sim1.6$), two very
metal-poor halo stars CS 22892-052 and CS 31082-001
([Fe/H]$\approx-3$) merit special attention. It is surprising that
the abundance patterns of the heavier neutron-capture elements in
these two ``main r-process stars" match the solar-system r-process
pattern very closely (e.g., \cite{cow99,tru02,wan06,sne08}).
However, their lighter neutron-capture elements ($37\leq Z\leq47$,
i.e., from Rb to Ag) are too deficient to agree entirely with the
solar-system's residual r-process pattern (e.g.,
\cite{sne00,hil02}). This implies that the r-process abundances of
solar-system for lighter neutron-capture elements, such as Sr, Y and
Zr, are not explained entirely by main r-process, but that another
process, referred to as the ``lighter element primary process"
(LEPP) \citep{tra04} or ``weak r-process" \citep{ish05}, is
required. By comparison, the abundance pattern of metal-poor stars
HD 122563 and HD 88609 is obviously different from the pattern
observed in CS 22892-052 and CS 31082-001. The abundances of HD
122563 and HD 88609 show excess of lighter neutron-capture elements
and deficiency of heavier neutron-capture elements
\citep{wes00,joh02,hon07}. Recently, the weak r-process (or LEPP)
abundances have been further investigated by many authors (e.g.,
\citet{mon07,izu09}). Based on spherically symmetric supernova
simulations, \cite{arc11} performed nucleosynthesis calculations and
found that LEPP (or weak r-process) abundances can be synthesized
under realistic conditions during the neutrino-wind phase.
Furthermore, heavier r-process elements can not be produced under
the conditions.

For the weak r-process stars HD 122563 and HD 88609, the ratios of
[Sr/Fe]$\approx0$ and [Eu/Fe]$\approx-0.5$ mean that weak r-process
elements are produced in conjunction with the Fe and light elements
rather than with heavier r-process elements. \citet{mon07} analyzed
the observed abundances of metal-poor stars with different [Sr/Eu]
ratios and concluded that the weak r-process produces a uniform and
unique abundance pattern of neutron-capture elements. In addition,
the analysis of \citet{qia07} show that the main r-process elements
are not produced in conjunction with light elements and iron group
elements. This means that the light elements and iron group elements
in the two main r-process stars (CS 22892-052 and CS 31082-001)
should have other astrophysical origins. Obviously, the observed
abundances of heavy elements in most metal-poor stars cannot be
explained by the abundances produced by a single nucleosynthetic
process. The yield ratios of ``high'' frequency (H) and ``low''
frequency (L) components have been obtained by \cite{qia07} from the
abundances of a main r-process star (CS 22892-052) and weak
r-process star (HD 122563), respectively. \cite{qia08} found the two
components were not enough to explain the phenomenon of a great
shortfall of Sr relative to Fe in many stars with
[Fe/H]$\lesssim-3$. They presented a third component related to
hypernovae (HNe) from progenitors of $\sim$25-50$M_\odot$.

A quantitative understanding of the astrophysical contributions of
neutron-capture elements to the Galaxy has so far been a challenging
problem \citep{sne08}. In order to better understand the origins of
neutron-capture elements in the Galaxy, one must separate main r-
and weak r-process contributions to the abundances observed in the
metal-poor stars. Although various models and methods have been
suggested to determine whether or not heavy element abundance
patterns in the stars are identical to the distributions in the
solar-system material, there is not yet a valid method to determine
the relative contributions from the individual process to the
lighter neutron-capture elements and heavier neutron-capture
elements. Obviously, more detailed studies of metal-poor stars are
needed to make progress in our understanding of the contributions
from individual neutron-capture process with various metallicities.
These reasons motivated us to start abundance study of the
metal-poor stars, in which light elements ($8\leq Z\leq20$, i.e.,
from O to Ca), iron group elements ($21\leq Z\leq30$, i.e., from Sc
to Zn), lighter neutron-capture elements ($38\leq Z\leq47$, i.e.,
from Sr to Ag) and heavier neutron-capture elements ($Z\geq56$,
i.e., heavier than Ba) are observed.

In this article, we fit observed abundances of 30 metal-poor stars
to derive the relative contributions from the individual r-process
to their elemental abundances. An abundance decomposition approach
of the stars is described in Section 2. The calculated results and
their physical meanings are presented in Section 3. Our conclusions
are given in Section 4.

\section{Abundance Distribution Approach of The Stars}

One of our major goals is to calculate the relative contribution of
the individual neutron-capture process to the elemental abundances
of metal-poor stars. We will start by exploring the origin of the
neutron-capture elements in the stars by comparing the observed
abundances with the predicted main r- and weak r-process
contributions. For this purpose, we propose that the abundance for
the $i$th element in a star can be calculated by the equation:

\begin{equation}
N_{i}([Fe/H])=(C_{r,m}N_{i,r,m}+C_{r,w}N_{i,r,w})\times10^{[Fe/H]},
\end{equation}
where $N_{i,r,m}$ and $N_{i,r,w}$ are the abundances of the $i$th
element produced by the main r-process and weak r-process,
respectively, which are normalized to [Fe/H]=0. $C_{r,m}$ and
$C_{r,w}$ are the corresponding component coefficients. Using
component coefficients, we can determine the relative contributions
of each process to the elemental abundances and then compare them
with the corresponding component coefficients of the solar system
(or [Fe/H]=0), in which $C_{r,m}=C_{r,w}=1$.

\subsection{Two r-process Components}

It is usually assumed that the weak r-process and main r-process
occur in explosive environments such as some regions inside
core-collapse SNe. There are two possible routes leading to a
core-collapse SN: one is collapse of an Fe core produced by
progenitors of $M\gtrsim11M_{\odot}$, in which both light elements
and iron group elements are produced; the other is collapse of an
O-Ne-Mg core produced by progenitors of $8-10M_{\odot}$, in which
the light elements and iron group elements are not produced. The
abundances in the weak r-process stars HD 122563 and HD 88609 are
important to constrain the weak r-process, since they have the
smallest contribution from the main r-process ([Eu/Fe]$\sim-0.5$)
and excess of the lighter neutron-capture elements Sr, Y, Zr. The
abundances of both lighter neutron-capture and light elements in
weak r-process stars should reveal the composition of the cloud
polluted mainly by the weak r-process event. For normal metal-poor
stars, the mean value of [Eu/Fe]$\approx0.3$ \citep{fie02} has been
identified, which means that the abundance of Eu is moderately
enhanced in the interstellar medium (ISM). In a cloud polluted by
the weak r-process event, the ratio of [Eu/Fe]$\approx-0.5$ for weak
r-process stars means that the ratio of [Eu/Fe] had changed from 0.3
to -0.5 due to pollution by weak r-process events, which implies
that the production of iron group and other light elements is
decoupled from that of heavier r-process elements, such as Eu.
However, [Sr/Fe]$\approx0$ observed in the weak r-process stars
implies that, although Fe and light elements (synthesized in
progenitor of SN II) are not produced by weak r-process
nucleosynthesis, they are ejected with the weak r-process elements
produced in the SN II. This also implies that, although light and
iron group elements are not produced by weak r-process
nucleosynthesis, both iron group elements and the weak r-process
elements in the weak r-process stars mainly come from a single
polluted event. So, the light elements, iron group elements and weak
r-process elements can be combined as one component. Because the
main r-process elements are not produced in conjunction with light
elements and iron group elements \citep{qia07}, the abundances of
main r-process stars (CS 22892-052 and CS 31082-001) could be
explained by stellar formation in a molecular cloud that had first
been polluted with weak r-process material, and then polluted by
main r-process material.

We take the average abundances in metal-poor stars HD 122563 and HD
88609 as a representative of the abundance distribution produced
mainly by the weak r-process. For the abundance distribution
dominated by the main r-process, we take the average abundances of
neutron-capture elements in metal-poor stars CS 22892-052 and CS
31082-001. In order to derive the pure weak r- and main
r-components, firstly, we can obtain the first order approximation
of the weak r-process abundance pattern by subtracting the average
abundances of CS 22892-052 and CS 31082-001 from the average
abundance of HD 122563 and HD 88609. To remove the effect of
metallicity and gas mixing during the star formation, the abundance
patterns are scaled by the average Eu abundance of HD 122563 and HD
88609.  Then, we derive the first order approximation of the main
r-process abundance pattern by subtracting the first order
approximation of the weak r-process abundance pattern from the
average abundance of CS 22892-052 and CS 31082-001, normalized to
Fe. Repeating above steps, we can derive the pure weak r- and pure
main r-components until the difference between the $n$th-order
approximation and the $(n-1)$th-order approximation are smaller than
the observational errors. This method has been simply described in
\cite{li12}. Note that, the weak r-process abundance pattern
includes light elements and iron group elements. Although they are
not produced by weak r-process nucleosynthesis, the inclusion of the
abundances of light and iron group elements in the weak r-component
means that the light and iron group elements (synthesized in
progenitor of SN II as primary-like yields) are ejected with the
weak r-process elements (produced in the SN II).

To compare main r-process and weak r-process abundance patterns
derived from the metal-poor stars with the r-process abundances of
solar-system, we use the main r-process and weak r-process abundance
patterns to fit solar-system r-process abundances. We adopt
solar-system r-process abundances from \citet{arl99} and
\citet{tra04}(for Sr-Nb). We find that the abundance of Ag cannot be
fitted. The solar r-process abundance of Ag \citep{arl99} is about
three times the sum of $N_{Ag;r;m}$ and $N_{Ag;r;w}$. Since Ag is
mainly an r-process element (about 80\%), one possible explanation
is the uncertainty of non-local thermodynamic equilibrium (NLTE)
effects in the abundance analysis of the metal-poor giant stars
\citep{han12}. The mixture of two r-process abundances matches well
the abundances of solar r-process (with the minimum $\chi^2$=0.56),
except for Ag. In this case, the main and weak r-process abundances
have been normalized to solar-system r-process abundances and the
normalized abundances of $N_{i,r,m}$ and $N_{i,r,w}$ in equation (1)
have been obtained. The abundances of $N_{i,r,m}$ and $N_{i,r,w}$
are listed in Table 1. It is noteworthy that there are some
discrepancies between the solar r-process abundances and the sum of
$N_{i,r,m}$ and $N_{i,r,w}$. For example, the solar r-process
abundance of Sr is about a factor of 0.83 smaller than the sum of
$N_{Sr,r,m}$ and $N_{Sr,r,w}$, but still within 1$\sigma$ of the
error bar. To determine the main r-process abundances and weak
r-process abundances, the observational abundances of metal-poor
stars are used. So, the observational uncertainties
(0.1$\sim$0.3dex) are contained in the derived two r-process
abundances. Furthermore, the determination of the solar r-process
abundances suffers from the uncertainties in predicting the
s-process contributions and the abundance uncertainties of the solar
system. Considering the uncertainties mentioned above, the
discrepancies between the sum of the two r-process abundances and
solar r-process abundances would be explained.

Because the solar r-process abundances can be matched by the sum of
$N_{i,r,m}$ and $N_{i,r,w}$, we can estimate the relative
contributions from the main and weak r-process components to the
abundances of neutron-capture elements in the solar system. We use
the ``percentage of weak r-process component" $f^{r}_{r,w}$ (i.e.,
$N_{i,r,w}/(N_{i,r,m}+N_{i,r,w})$) and the ``percentage of main
r-process component" $f^{r}_{r,m}$ (i.e.,
$N_{i,r,m}/(N_{i,r,m}+N_{i,r,w})$) to estimate the relative
contributions to r-process abundances of solar system. In Fig. 1, we
plot $f^{r}_{r,w}$ and $f^{r}_{r,m}$ as functions of atomic number
Z. We find that there is a decreasing trend in $f^{r}_{r,w}$ and an
increasing trend in $f^{r}_{r,m}$ as atomic number increase from
Z=30 to Z=63, which are close to linearity. The fractions of
contribution to solar system from primary-like yields (or ``weak
r-component") $f_{r,w}$ ($f_{r,w}$=$N_{i,r,w}/N^{total}_{\odot}$)
are listed in the fifth column of Table 1. The fraction of
contribution from primary-like yields to Fe is about 25\%. The
abundances of solar system is given by \cite{and89}. Recently,
\cite{lo09} updated the abundances of solar system, the new values
should not change our results and conclusions.

\subsection{Abundance Fit Approach}

Nearly all chemical evolution and nucleosynthetic information is in
the form of elemental abundances in stars with various
metallicities. In order to investigate the relative contributions
from the individual neutron-capture process, the abundances in the
metal-poor halo stars are particularly useful. In this work we
analyze the direct observational constraints provided by the
photospheric composition of the metal-poor stars. Our goal is to
find the parameters that characterize the observed data.  The
standard definitions of elemental abundances and ratios are used
throughout the paper. For element X, the abundance is defined as the
logarithm of the number of atoms of element X per $10^{12}$ hydrogen
atoms, log$\varepsilon(X)\equiv$log$_{10}(N_{X}/N_{H})$+12.0. The
abundance ratio relative to the solar ratio of element X and element
Y is defined as
[X/Y]$\equiv$log$(N_{X}/N_{Y})$-log$(N_{X}/N_{Y})_{\odot}$. The
reduced $\chi^2$ is defined as

\begin{equation}
\chi^2=\sum_{i=1}^K\frac{(logN_{i,obs}-logN_{i,cal})^2}{(\Delta
logN_{i,obs})^2(K-K_{free})},
\end{equation}
where $logN_{i,obs}$ is the observed abundance of the $i$th element,
$\Delta logN_{i,obs}$ is the observational error, $N_{i,cal}$ is the
calculated abundance from equation (1), $K$ is the number of
elements applied in the fit and $K_{free}$ is the number of free
parameters. We use the parametric approach to investigate what the
possible relative contributions might be to reproduce the observed
abundance patterns found in metal-poor stars. Based on equation (1),
we carry out the calculation, including the contributions of the
main r-process and weak r-process to fit the abundances observed in
the stars, in order to look for the minimum $\chi^2$ in the
two-parameter space formed by $C_{r,m}$ and $C_{r,w}$. For a good
fit, the reduced $\chi^2$ should be of order unity.

\section{Results and Discussions}

\subsection{Fitted Results}

It is noted that the weak r-process component in equation (1)
include light elements (from O to Ca), iron group elements (from Sc
to Zn), lighter neutron-capture elements (such as Sr, Y and Zr) and
heavier neutron-capture elements (such as Ba), so the observed data
of investigated sample stars should contain abundances of these four
groups of elements. In order to obtain better constraint on main
r-process component coefficient, these observed data must contain
the abundances of Eu and the number of observed neutron-capture
elements should be more than four. We select 30 metal-poor stars
([Fe/H]$\leq-2.0$) with [Ba/Fe]$<0.5$, which means that the
s-enhanced stars are not included. The parameters can be obtained
using the observed data in 30 sample stars collected by
\citet{sud08} and from literatures
\citep{wes00,cow02,hil02,joh02,joh02b,sne03,hon04,hon06,hon07,bar05,iva06,chr08,hay09,mas10,roe10}.
The results of the component coefficients, $\chi^2$ and
($K-K_{free}$) are listed in Table 2.

Two examples are provided in Figs. 2 and 3, which shows our
calculated results for two sample stars, HE 1319-0312 and HD 186478.
In order to provide a convenient comparison, the observed elemental
abundances are marked by filled circles. The dotted line represents
the solar-system r-process abundances given by \cite{arl99} (from Sr
to Nb, the abundances are updated by \cite{tra04} and the abundances
of Pb and Th come from \cite{sne08}), which is normalized to Eu. The
predictions are most within the observational error for elements
from O to Pb. In the top panel of Fig. 4, we show individual
relative offsets ($\Delta\log\varepsilon$) for the sample stars,
except for BD-18 5550, between the predictions with the observed
abundances. Typical observational uncertainties in $\log\varepsilon$
presented by dotted lines are $\sim0.2-0.3$ dex. The rms offsets of
these elements in $\log\varepsilon$ are mostly smaller than 0.20
dex, which are shown in the bottom panel. These values are within
the uncertainties of abundance determinations. From Fig. 4 we can
see that the predictions are in agreement with the observed
abundances for most sample stars, not only for light elements and
iron group elements but also for the lighter neutron-capture
elements and the heavier neutron-capture elements. The good
agreement supports the idea that there exist two robust processes.

\subsection{The Trends of Component Coefficients and The Logarithmic Component Ratios}

We can obtain some information from the component coefficients.
$C_k>1$ (or $C_k<1$, $k=r,m$ or $r,w$) means that, except for the
effect of metallicity, the contribution from the corresponding
process to the neutron-capture-element abundances in the sample star
is larger (or less) than that in the solar system. If two
coefficients are not equal to each other, the relative contributions
from the various components to the neutron-capture-element
abundances are not in proportion to that of the solar-system. The
component coefficients as a function of metallicity, illustrated in
Fig. 5, contain some important information. For the metal-poor
stars, since they are thought to exhibit an abundance pattern
produced by a few r-process events in the early Galaxy, the very
large scatter of $C_{r,m}$ from star to star samples a largely
unmixed early Galaxy. This unmixed behavior had been studied by many
authors \citep{tra01,fie02,joh02b,cow05,ces08}. We note that for
most metal-poor stars, both $C_{r,w}$ and $C_{r,m}$ are larger than
1. This means that the contributions from the weak r- and main
r-processes to the neutron-capture-element abundances in these
stars, except for the effects of metallicity, are larger than those
in the solar system. This result can be naturally explained by the
shorter lifetimes of massive stars: massive stars evolve quickly,
ending as SNe II producing r-process elements. As the metallicity
increases, a very large scatter of main r-process component
coefficients is obtained for metal-poor stars for the range of
$-3\lesssim$[Fe/H]$\lesssim-2.5$, then the scatter begin to
decrease.

In order to study Galactic evolution of neutron-capture elements,
the variation of the logarithmic ratio [element/Fe] with metallicity
is particularly useful \citep{tra99}. For the $i$th element, the
logarithmic component ratio of the individual neutron-capture
process relative to solar ratio [element/Fe]$_k$ ($k=r,w; r,m;$) is
defined as:

\begin{equation}
[element/Fe]_{k}=log(C_{k}N_{i,k}\times10^{[Fe/H]})-logN_{Fe}-(logN_{i}-logN_{Fe})_{\odot}=log(C_{k}N_{i,k})-logN_{i,\odot},
\end{equation}
where $N_{i,k}$ is the abundances of the $i$th element produced by
the individual neutron-capture process and $N_{i,\odot}$ is
abundance of the element in the solar system. The derived component
ratios are shown in Fig. 6, in which Sr, Y, Zr and Ba are taken as
examples. The trend of ratios [element/Fe]$_{r,w}$ is almost
constant for the normal metal-poor stars, which is clearly different
from those of [element/Fe]$_{r,m}$. We find that the ratios
[Sr/Fe]$_{r,w}$ are nearly -0.3 except for BD-18 5550.

\subsection{The Application of Approach: The Finding of Specific
Objects}

Using the component coefficients, we can select those stars with
special neutron-capture-element abundance distributions. If one
component coefficient is much larger than others, this star might
have been formed in a Galactic region that was not well-mixed
chemically and the corresponding process may be dominantly
responsible for the neutron-capture elements in this star. For
example, the r-process component coefficients of metal-poor star CS
29497-004 are $C_{r,m}$ =73.16 and $C_{r,w}$ =3.73. This implies
that CS 29497-004 is also a main r-process star.

There are a number of studies that have been done to explore the
origin of neutron-capture elements over many years. Therefore, the
observed abundances of metal-poor stars provide the best opportunity
to investigate the abundance patterns produced by individual
neutron-capture processes \citep{sne08}. The detailed abundance
studies for main r-process stars have revealed that the abundance
pattern of heavier neutron-capture elements ($Z\geq56$) is close to
the r-process pattern of solar system. However, their lighter
neutron-capture elements are deficient to the solar r-process
pattern (e.g., \cite{sne00,hil02}), which implies that another
component, such as weak r-process component, is required. The
metal-poor stars HD 122563 and HD 88609 have extreme excesses of
lighter neutron-capture elements and are considered as weak
r-process stars. Clearly, it is very important to find more weak
r-process stars to investigate the robustness of weak r-process
pattern.

Based on our results shown in Fig. 5 and Table 2, we can find that
the metal-poor stars, BD+4 2621 with $C_{r,m}$=0.36, $C_{r,w}$=3.85
and HD 4306 with $C_{r,m}$=0.33, $C_{r,w}$=5.79, could be two weak
r-process stars. In order to test this finding, Fig. 7 shows the
abundance comparisons on the logarithmic scale among BD+4 2621, HD
4306 and HD 122563 as a function of atomic number, which have been
normalized to Fe abundance of HD 122563. As a comparison, the
abundances of HD 88609 are also reported. Fig. 7 reveals that the
abundance patterns of the two stars (BD+4 2621 and HD 4306) are
quite similar to the abundance pattern of HD 122563, although the
differences do not seem significant. Our conclusion here is that
BD+4 2621 and HD 4306 are another two examples of weak r-process
stars. Please note that BD-18 5550 is not a weak r-process star, we
will discuss it in Section 3.4. The metallicity of BD+4 2621
([Fe/H]=-2.52) is even higher than that of HD 122563. This means
that the weak r-process stars, HD 122563 and HD 88609, are not
peculiar objects and the abundance pattern produced by weak
r-process is remarkably stable from star to star, at least in the
range of metallicity between -2.52 and -3.1.

One could recall that the abundance patterns of the heavier elements
for two main r-process stars match the solar-system r-process
pattern very closely and their lighter neutron-capture elements
($37\leq Z \leq 47$, i.e., from Rb to Ag) are too deficient to agree
entirely with the solar-system's residual r-process pattern. This
phenomenon is also shown in right panel of Fig. 2 for r-rich star HE
1219-0312. From the component coefficients listed in Table 2, we
find the main r-process coefficient is close to the weak r-component
for some sample stars (e.g., HD 6268, HD 186478 and HD 165195). For
example, the main r-process coefficient and the weak r-component
coefficient of metal-poor star HD 186478 are 4.42 and 4.28,
respectively. As shown in right panel of Fig. 3, the abundances of
the neutron-capture elements are well matched by the scaled solar
r-process pattern. Obviously, the abundances of HD 186478 also seem
to be a better match to the scaled r-process abundances in solar
system for lighter neutron-capture elements Sr, Y and Zr. The
similar values of two r-process coefficients imply that the relative
contributions from two components to the neutron-capture-element
abundances are in proportion to that of the solar system.

\subsection{Investigation of The Prompt Component}

Based on observations, there are some metal-poor stars with
[Sr/Fe]$\lesssim-1$ (see Fig. 4 in \citet{tra04}). In our sample
stars, there is only one star, BD-18 5550, with [Sr/Fe]$<-1$. From
Table 2, we can see that its $\chi^{2}$ is larger than 15.
Obviously, the observed abundances of metal-poor stars with
[Sr/Fe]$\lesssim-1$ can not be fitted by weak r-process and main
r-process components. So there should be another component which
mainly produces iron group elements and light elements, and barely
produces neutron-capture elements in the early Galaxy. This
component was called as the initial or prompt (P) component by
\cite{qia01}. The P component may arise from the first generations
of very massive stars. These stars produce the initial abundance of
Fe and associated elements corresponding to [Fe/H]$\lesssim-3$, and
the observed abundance of metal-poor stars with [Sr/Fe]$\lesssim-1$
should be the representative of P component. The low ratios of
[Sr/Fe]$\lesssim-1$ (the lowest observed values of [Sr/Fe] are about
-2.0 for metal-poor stars) imply that P component mainly produces
iron group elements and light elements, and barely produce
neutron-capture elements.

The abundances of BD-18 5550 ([Fe/H]=-3.05 and [Sr/Fe]=-1.2) have
been observed by \cite{joh02} and \cite{joh02b}. We adopt the main
r-process and weak r-process abundance patterns to fit the
abundances of neutron-capture elements in BD-18 5550. The fitted
result is shown in Fig. 8 by solid line. The components coefficients
and $\chi^2$ deduced for this star are $C_{r,m}$ =0.71, $C_{r,w}$
=0.51 and $\chi^2=1.19$. Although the star is the low-[Sr/Fe] star,
its abundances of neutron-capture elements can be fitted by two
r-process components. From Fig. 8 we can see that, for the light
elements and iron group elements, the observed abundances are higher
than calculated result, because the contribution of the P component
is not included. We can estimate the P component by subtracting
calculated values from observed abundances for light elements and
iron group elements; the abundance pattern of P component is listed
in Table 3. Note that, once the SNe II in which neutron-capture
elements are produced began to contribute their production to ISM,
the effect of P component to the stellar abundances become smaller
\citep{qia01}.

As a test, we have used three components (main r-, weak r- and
P-components) to fit the abundances of sample stars. In this case,
the abundance for the $i$th element in a star can be calculated as:

\begin{equation}
N_{i}([Fe/H])=(C_{r,m}N_{i,r,m}+C_{r,w}N_{i,r,w}+C_{P}N_{i,P}/10^{[Fe/H]_{185550}})\times10^{[Fe/H]},
\end{equation}
where [Fe/H]$_{18 5550}$=-3.05, $N_{i,P}$ is the abundance of the
$i$th element produced by the P-component (listed in Table 3) and
$C_{P}$ is the P-component coefficient. The component coefficients
for main r-, weak r- and P-components are listed in Table 4 and
their trends are shown in Fig. 9. From Fig. 9 we can see that the
coefficients of P-component decrease with increasing metallicities.
Since the abundances of the metal-poor halo stars can be used as a
probe of the conditions that existed in the history of the Galaxy,
it is important to determine the relative contributions from the
individual process to the elemental abundances in the stars. Based
on equation (4), we can isolate the contributions corresponding to
the primary-like yields (or weak r-component) and P-component to
light elements and iron group elements. The component fractions
$f_{i,P}$ can be calculated as:

\begin{equation}
f_{i,P}=\frac{C_{P}N_{i,P}\times10^{[Fe/H]-[Fe/H]_{18
5550}}}{N_{i}},
\end{equation}
where $N_{i}$ is calculated from equation (4). In Fig. 10, the
component fractions $f_{i,P}$ for element Fe as function of [Fe/H]
are given. In our sample stars, BD-18 5550 has the lowest ratio of
[Sr/Fe] ([Sr/Fe]=-1.2) and the largest component fractions
$f_{Fe,P}$ ($f_{Fe,P}$=0.88). Except for BD-18 5550, the abundances
of star BD-17 6036 with [Sr/Fe]=-0.61 also apparently contains the
contribution from P-component. There is an upper limit of component
fractions $f_{Fe,P}$ at a given [Fe/H]. We can see that the upper
limit of $f_{Fe,P}$ decrease linearly with increasing metallicities.
At metallicity [Fe/H]$\sim-2.0$, the contributions from P-component
are smaller than about 25\%. This means that the contribution of
P-component would be invisible for higher metallicities
([Fe/H]$\gtrsim-1.5$).

\section{Conclusions}

In this work, we present an approach to fit the abundances of
metal-poor stars and determine the relative contributions from
individual neutron-capture process. Our results can be summarized as
follows:

1. The abundances of neutron-capture elements in all metal-poor
stars, including main r-process stars and weak r-process stars,
contain the contributions of two r-processes. For weak r-process
stars, $C_{r,w}\gg C_{r,m}$; for main r-process stars, $C_{r,m}\gg
C_{r,w}$.

2. The weak r-process coefficients are close to constant for normal
metal-poor stars, including r-rich stars. The component ratio
[Sr,Y,Zr/Fe]$_{r;w}$ is also nearly constant for most metal-poor
stars. Our conclusion is that the elements produced by weak
r-process have increased along with Fe over the polluted history.

3. For most sample stars, the relative contribution from the
individual neutron-capture process to the heavy element abundances
was not usually found to be in the solar proportion, especially for
main r-process stars and weak r-process stars. However, we find some
stars whose main r-process coefficients are close to their weak
r-process coefficients. The abundance pattern of the neutron-capture
elements in these stars is close to the solar r-process pattern.

4. In addition to the well-known weak r-process stars HD 122563
([Fe/H]=-2.77) and HD 88609 ([Fe/H]=-3.07), based on the component
coefficients, we find that the metal-poor stars BD+4 2621 and HD
4306 are also weak r-process stars. The metallicity of BD+4 2621 is
[Fe/H]=-2.52, which means that abundance pattern produced by weak
r-process is stable from star to star, at least in the range of
metallicity between -2.52 and -3.1.

5. In our sample stars, there is only one star, BD-18 5550, having
[Sr/Fe]$<-1$. Although the star is the low-[Sr/Fe] star, its
abundances of neutron-capture elements can also be fitted by two
r-process components. However, the observed abundances for the light
elements and iron group elements are higher than estimated values,
because the contribution of the P-component is not included. The
abundance pattern of P component is estimated from BD-18 5550. There
is an upper limit of fractions of P-component at a given [Fe/H]. We
find that the upper limit of the fraction decrease linearly with
increasing metallicities. This means that the contribution of
P-component would be invisible for higher metallicities
([Fe/H]$\gtrsim-1.5$).

6. It is very important to note that the solar-system r-process
abundances can be best-fitted by main r-process and weak r-process
abundance patterns derived from the very metal-poor stars, which
implies that both r-process abundance patterns are weakly
dependent on metallicity. This also means that solar r-process
abundances have been quantitatively decomposed into two
components. In solar system, the trend of percentage of weak
r-process component linearly decrease with atomic number from Zn
to Eu; that of main r-process component is contrary. The fraction
of contribution from primary-like yields to Fe in solar system is
estimated about 25\%.

Our results could give the constraints on models of the r- and
s-processes that yield lighter and heavier neutron-capture elements
in the Galaxy. Our hope is that the results here will provide useful
information to explore the origins of neutron-capture elements in
the Galaxy. Obviously, more detailed abundances of neutron-capture
elements and light elements in more stars are needed.

\acknowledgments

We thank the referee for very valuable comments and suggestions that
have improved this article greatly. This work has been supported by
the National Natural Science Foundation of China under Grant No.
11273011, U1231119, 10973006 and 11003002, the Science Foundation of
Hebei Normal University under Grant No. L2009Z04, the Natural
Science Foundation of Hebei Province under Grant No. A2009000251,
A2011205102, Science and Technology Supporting Project of Hebei
Province under Grant No. 12211013D and the Program for Excellent
Innovative Talents in University of Hebei Province under Grant No.
CPRC034.

\clearpage

\begin{table}
\begin{center}
\caption{The two r-process component and the fractions of
contribution to solar system from primary-like yields (or ``weak
r-component").\label{tbl-1}}
\begin{tabular*}{458pt}{@{\extracolsep\fill}lcccc}
\tableline\tableline
Element &  Z  & $N_{i,r,m}$ & $N_{i,r,w}$ &  $f_{r,w}$\\
\tableline
O   &   8   &   0.00E+00    &   1.97E+07    &   0.83    \\
Na  &   11  &   0.00E+00    &   1.33E+04    &   0.23    \\
Mg  &   12  &   0.00E+00    &   7.09E+05    &   0.66    \\
Al  &   13  &   0.00E+00    &   5.56E+03    &   0.07    \\
Si  &   14  &   0.00E+00    &   6.49E+05    &   0.65    \\
Ca  &   20  &   0.00E+00    &   3.57E+04    &   0.58    \\
Sc  &   21  &   0.00E+00    &   1.01E+01    &   0.30    \\
Ti  &   22  &   0.00E+00    &   1.28E+03    &   0.53    \\
V   &   23  &   0.00E+00    &   9.37E+01    &   0.32    \\
Cr  &   24  &   0.00E+00    &   3.13E+03    &   0.23    \\
Mn  &   25  &   0.00E+00    &   8.19E+02    &   0.09    \\
Fe  &   26  &   0.00E+00    &   2.26E+05    &   0.25    \\
Co  &   27  &   0.00E+00    &   8.65E+02    &   0.38    \\
Ni  &   28  &   0.00E+00    &   9.57E+03    &   0.19    \\
Cu  &   29  &   0.00E+00    &   2.28E+01    &   0.04    \\
Zn  &   30  &   0.00E+00    &   4.57E+02    &   0.36    \\
 Sr     &   38  &   1.92E+00    &   3.74E+00    &   0.16    \\
 Y  &   39  &   2.18E-01    &   6.10E-01    &   0.13    \\
 Zr     &   40  &   1.11E+00    &   3.06E+00    &   0.27    \\
 Nb     &   41  &   1.09E-01    &   1.33E-01    &   0.19    \\
Mo  &   42  &   2.42E-01    &   6.35E-01    &   0.25    \\
 Ru     &   44  &   8.85E-01    &   6.53E-01    &   0.35    \\
 Rh     &   45  &   1.60E-01    &   3.23E-01    &   0.94    \\
 Pd     &   46  &   3.63E-01    &   2.30E-01    &   0.17    \\
 Ag     &   47  &   7.75E-02    &   5.52E-02    &   0.11    \\
\tableline
\end{tabular*}
\end{center}
\end{table}

\begin{table}
\begin{center}
{Table 1 continued.\label{tbl-1}}
\begin{tabular*}{458pt}{@{\extracolsep\fill}lcccc}
\tableline\tableline
Element &  Z  & $N_{i,r,m}$ & $N_{i,r,w}$ &  $f_{r,w}$\\
\tableline
 Ba     &   56  &   9.26E-01    &   2.46E-02    &   0.01    \\
 La     &   57  &   1.05E-01    &   1.32E-03    &   0.00    \\
 Ce     &   58  &   2.19E-01    &   3.93E-02    &   0.03    \\
 Pr     &   59  &   5.71E-02    &   3.23E-02    &   0.19    \\
 Nd     &   60  &   3.16E-01    &   1.73E-02    &   0.02    \\
 Sm     &   62  &   1.56E-01    &   1.39E-02    &   0.05    \\
 Eu     &   63  &   8.24E-02    &   0.00E+00    &   0.00    \\
 Gd     &   64  &   2.56E-01    &   0.00E+00    &   0.00    \\
Tb  &   65  &   3.91E-02    &   0.00E+00    &   0.00    \\
 Dy     &   66  &   3.50E-01    &   0.00E+00    &   0.00    \\
Ho  &   67  &   8.83E-02    &   0.00E+00    &   0.00    \\
 Er     &   68  &   2.41E-01    &   0.00E+00    &   0.00    \\
 Tm     &   69  &   2.81E-02    &   0.00E+00    &   0.00    \\
Yb  &   70  &   1.84E-01    &   0.00E+00    &   0.00    \\
Lu  &   71  &   4.97E-02    &   0.00E+00    &   0.00    \\
Hf  &   72  &   8.96E-02    &   0.00E+00    &   0.00    \\
Os  &   76  &   1.00E+00    &   0.00E+00    &   0.00    \\
 Ir     &   77  &   7.10E-01    &   0.00E+00    &   0.00    \\
Pt  &   78  &   1.16E+00    &   0.00E+00    &   0.00    \\
Au  &   79  &   9.25E-02    &   0.00E+00    &   0.00    \\
Pb  &   82  &   6.85E-01    &   0.00E+00    &   0.00    \\
 Th     &   90  &   3.25E-02    &   0.00E+00    &   0.00    \\
U   &   92  &   5.20E-03    &   0.00E+00    &   0.00    \\

\tableline
\end{tabular*}
\tablecomments{$\log\varepsilon$=log$N$+1.54.}
\end{center}
\end{table}

\begin{table}
\begin{center}
\caption{The results of the two component coefficients, $\chi^{2}$
and $K-K_{free}$ for sample stars.\label{tbl-2}}
\begin{tabular*}{458pt}{@{\extracolsep\fill}lccccc}
\tableline\tableline
Star &  [Fe/H]  & $C_{r,m}$ & $C_{r,w}$ &  $\chi^2$  & $K-K_{free}$\\
\tableline
HD 221170   &   -2.18   &   6.14    &   3.91    &   1.23    &   41  \\
HE 1219-0312    &   -2.96   &   31.45   &   2.78    &   0.78    &   31  \\
CS 31082-001    &   -2.91   &   52.04   &   3.89    &   0.64    &   37  \\
CS 29497-004    &   -2.63   &   73.16   &   3.73    &   0.97    &   15  \\
CS 29491-069    &   -2.51   &   13.80   &   3.57    &   1.10    &   26  \\
HD 115444   &   -2.98   &   7.98    &   4.90    &   1.70    &   36  \\
BD +17$^{o}$03248 &   -2.08   &   9.10    &   3.87    &   0.70    &   42  \\
CS 22892-052    &   -3.10   &   50.31   &   3.61    &   0.92    &   44  \\
HD 6268 &   -2.63   &   4.16    &   4.02    &   0.96    &   28  \\
HD 122563   &   -2.77   &   0.26    &   4.07    &   1.08    &   34  \\
HD 88609 &   -3.07   &   0.56    &   4.73    &   1.93    &   30  \\
HE 2224+0143    &   -2.58   &   15.69   &   3.77    &   0.39    &   16  \\
HE 2252-4225    &   -2.82   &   13.47   &   4.07    &   0.69    &   19  \\
HE 2327-5642    &   -2.78   &   12.39   &   2.04    &   1.29    &   35  \\
HD 108577   &   -2.38   &   2.52    &   4.46    &   1.07    &   25  \\
HD 165195   &   -2.32   &   3.22    &   3.39    &   1.30    &   20  \\
HD 216143   &   -2.23   &   3.04    &   4.02    &   0.56    &   21  \\
BD-18 5550  &   -3.05   &   0.56    &   3.15    &   15.07    &   20  \\
BD-17 6036  &   -2.77   &   1.30    &   3.92    &   3.24    &   21  \\
BD-11 145   &   -2.50   &   3.27    &   4.18    &   1.13    &   18  \\
BD+4 2621   &   -2.52   &   0.36    &   3.85    &   0.73    &   18  \\
BD+5 3098   &   -2.74   &   1.98    &   4.45    &   1.42    &   20  \\
BD+8 2856   &   -2.12   &   2.87    &   3.76    &   1.36    &   27  \\
BD+9 3223   &   -2.29   &   3.93    &   4.58    &   1.33    &   18  \\
BD+10 2495  &   -2.08   &   3.14    &   3.47    &   0.55    &   19  \\
\tableline
\end{tabular*}
\end{center}
\end{table}

\begin{table}
\begin{center}
{Table 2 continued.\label{tbl-2}}
\begin{tabular*}{458pt}{@{\extracolsep\fill}lccccc}
\tableline\tableline
Star &  [Fe/H]  & $C_{r,m}$ & $C_{r,w}$ &  $\chi^2$  & $K-K_{free}$\\
\tableline
HD 4306 &   -2.89   &   0.33    &   5.79    &   1.40    &   16  \\
HD 110184   &   -2.52   &   1.02    &   3.52    &   2.18    &   27  \\
HD 126587   &   -2.78   &   4.30    &   5.29    &   1.36    &   22  \\
HD 186478   &   -2.50   &   4.42    &   4.28    &   0.64    &   28  \\
CS 30306-132    &   -2.42   &   10.17   &   4.63    &   1.22    &   26  \\
\tableline
\end{tabular*}
\end{center}
\end{table}

\begin{table}
\begin{center}
\caption{The abundance pattern of P component estimated from BD-18
5550.\label{tbl-3}}
\begin{tabular*}{458pt}{@{\extracolsep\fill}cccccc}
\tableline\tableline
Element&Z&$N_{i,P}$&Element&$Z$&$N_{i,P}$\\
\tableline
Na  &   11  &   7.91E+01    &   V   &   23  &   1.76E-01    \\
Mg  &   12  &   3.94E+03    &   Cr  &   24  &   6.34E+00    \\
Al  &   13  &   2.26E+01    &   Mn  &   25  &   2.72E+00    \\
Si  &   14  &   2.66E+03    &   Fe  &   26  &   7.48E+02    \\
Ca  &   20  &   1.39E+02    &   Co  &   27  &   3.41E+00    \\
Sc  &   21  &   2.42E-02    &   Ni  &   28  &   4.94E+01    \\
Ti  &   22  &   2.65E+00    &   Zn  &   30  &   2.30E+00    \\
\tableline
\end{tabular*}
\tablecomments{$\log\varepsilon$=log$N$+1.54.}
\end{center}
\end{table}

\begin{table}
\begin{center}
\caption{The results of the three component coefficients, $\chi^{2}$
and $K-K_{free}$ for sample stars.\label{tbl-4}}
\begin{tabular*}{458pt}{@{\extracolsep\fill}lcccccc}
\tableline\tableline
Star &  [Fe/H]  & $C_{r,m}$ & $C_{r,w}$ & $C_{P}$ &  $\chi^2$  & $K-K_{free}$\\
\tableline
HD 221170   &   -2.18   &   6.12    &   4.05    &   0.00    &   1.11    &   38  \\
HE 1219-0312    &   -2.96   &   31.72   &   2.02    &   0.19    &   0.79    &   29  \\
CS 31082-001    &   -2.91   &   52.32   &   3.23    &   0.15    &   0.66    &   35  \\
CS 29497-004    &   -2.63   &   73.16   &   3.73    &   0.00    &   1.04    &   14  \\
CS 29491-069    &   -2.51   &   13.85   &   3.36    &   0.05    &   1.14    &   25  \\
HD 115444   &   -2.98   &   7.98    &   4.91    &   0.00    &   1.83    &   33  \\
BD +1703248 &   -2.08   &   9.14    &   3.69    &   0.06    &   0.74    &   39  \\
CS 22892-052    &   -3.1    &   50.44   &   2.74    &   0.21    &   0.87    &   41  \\
HD 6268 &   -2.63   &   4.17    &   3.86    &   0.06    &   1.00    &   27  \\
HD 122563   &   -2.77   &   0.26    &   3.63    &   0.16    &   1.08    &   31  \\
HD 88609    &   -3.07   &   0.41    &   4.89    &   0.00    &   1.47    &   28  \\
HE 2224+0143    &   -2.58   &   15.69   &   3.77    &   0.00    &   0.42    &   15  \\
HE 2252-4225    &   -2.82   &   13.82   &   3.26    &   0.21    &   0.71    &   18  \\
HE 2327-5642    &   -2.78   &   12.51   &   1.6 &   0.11    &   1.30    &   34  \\
HD 108577   &   -2.38   &   2.66    &   2.24    &   0.62    &   0.44    &   24  \\
HD 165195   &   -2.32   &   3.22    &   3.39    &   0.00    &   1.37    &   19  \\
HD 216143   &   -2.23   &   3.04    &   3.98    &   0.01    &   0.59    &   20  \\
BD-18 5550  &   -3.05   &   0.71    &   0.51    &   1.00    &   0.42    &   20  \\
BD-17 6036  &   -2.77   &   1.46    &   1.39    &   0.78    &   1.31    &   20  \\
BD-11 145   &   -2.5    &   3.77    &   2.14    &   0.51    &   0.57    &   17  \\
BD+4 2621   &   -2.52   &   0.37    &   3.16    &   0.26    &   0.63    &   17  \\
BD+5 3098   &   -2.74   &   2.06    &   2.52    &   0.52    &   1.03    &   19  \\
BD+8 2856   &   -2.12   &   2.9 &   3.14    &   0.15    &   1.37    &   26  \\
BD+9 3223   &   -2.29   &   4.21    &   3.08    &   0.37    &   1.16    &   17  \\
BD+10 2495  &   -2.08   &   3.28    &   2.57    &   0.22    &   0.47    &   18  \\
\tableline
\end{tabular*}
\end{center}
\end{table}

\begin{table}
\begin{center}
{Table 4 continued.\label{tbl-4}}
\begin{tabular*}{458pt}{@{\extracolsep\fill}lcccccc}
\tableline\tableline
Star &  [Fe/H]  & $C_{r,m}$ & $C_{r,w}$ & $C_{P}$ &  $\chi^2$  & $K-K_{free}$\\
\tableline
HD 4306 &   -2.89   &   0.33    &   5.31    &   0.12    &   1.45    &   15  \\
HD 110184   &   -2.52   &   1.02    &   3.48    &   0.00    &   2.33    &   25  \\
HD 126587   &   -2.78   &   4.3 &   5.29    &   0.00    &   1.43    &   21  \\
HD 186478   &   -2.5    &   4.42    &   4.28    &   0.00    &   0.69    &   26  \\
CS 30306-132    &   -2.42   &   10.17   &   4.63    &   0.00    &   1.27    &   25  \\

\tableline
\end{tabular*}
\end{center}
\end{table}

\clearpage

\begin{figure}[t]
 \centering
 \includegraphics[width=1\textwidth,height=0.6\textheight]{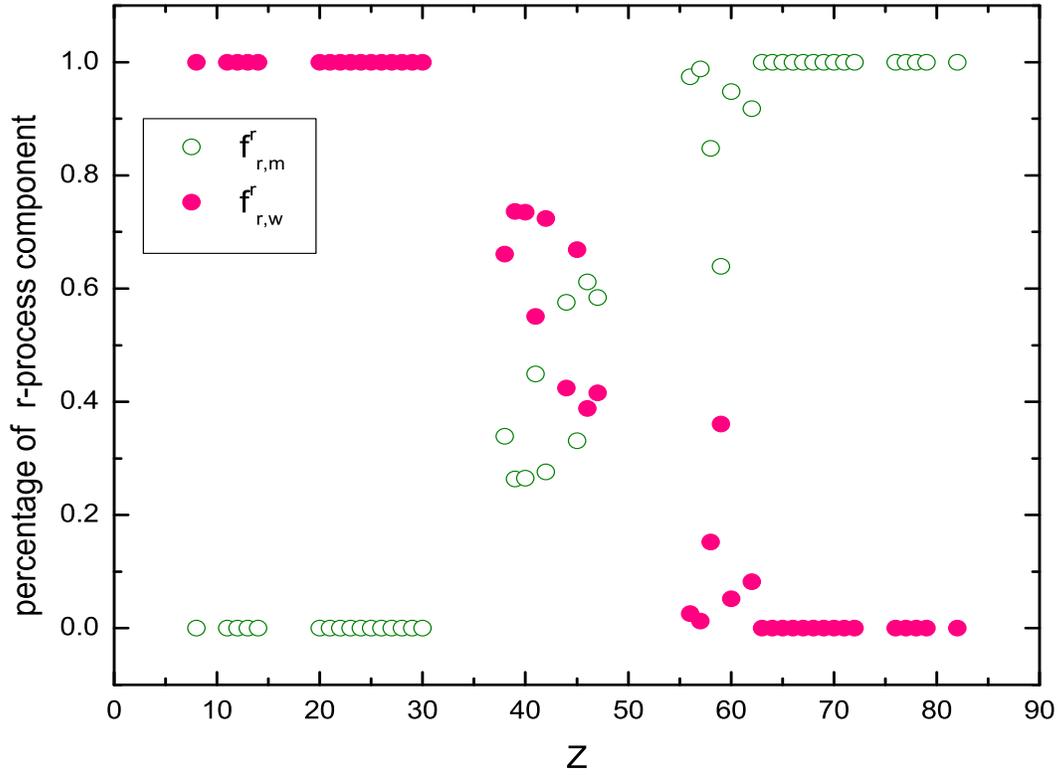}
\caption{Percentage of r-process components for elements. Symbols:
open circles and filled circles are percentage of main r-process
component and percentage of weak r-process component, respectively.}
\end{figure}

\begin{figure}[t]
 \centering
 \includegraphics[width=1\textwidth,height=0.5\textheight]{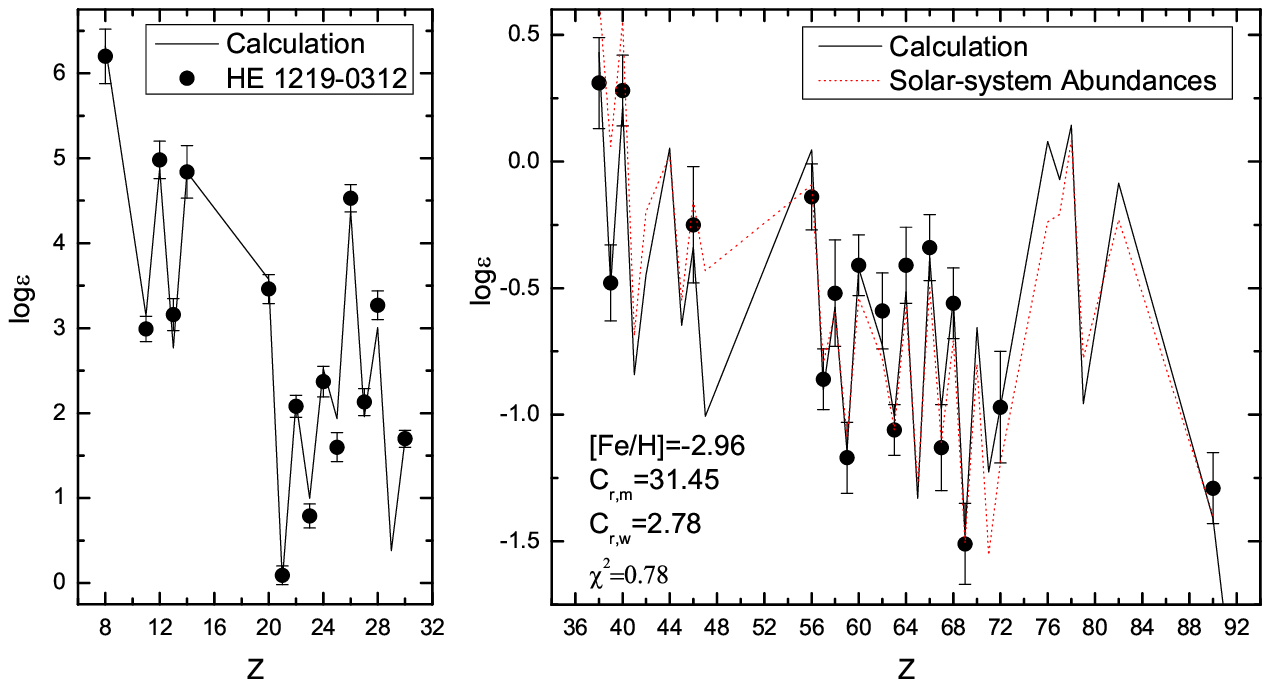}
\caption{Fitted result for the sample star HE 1219-0312. Symbols:
filled circles are observed abundances; the solid line presents the
calculated results and the dotted line is the scaled solar-system
r-process abundances. The left panel is the result for light
elements. The right panel is the result for neutron-capture
elements.}
\end{figure}

\begin{figure}[t]
 \centering
 \includegraphics[width=1\textwidth,height=0.5\textheight]{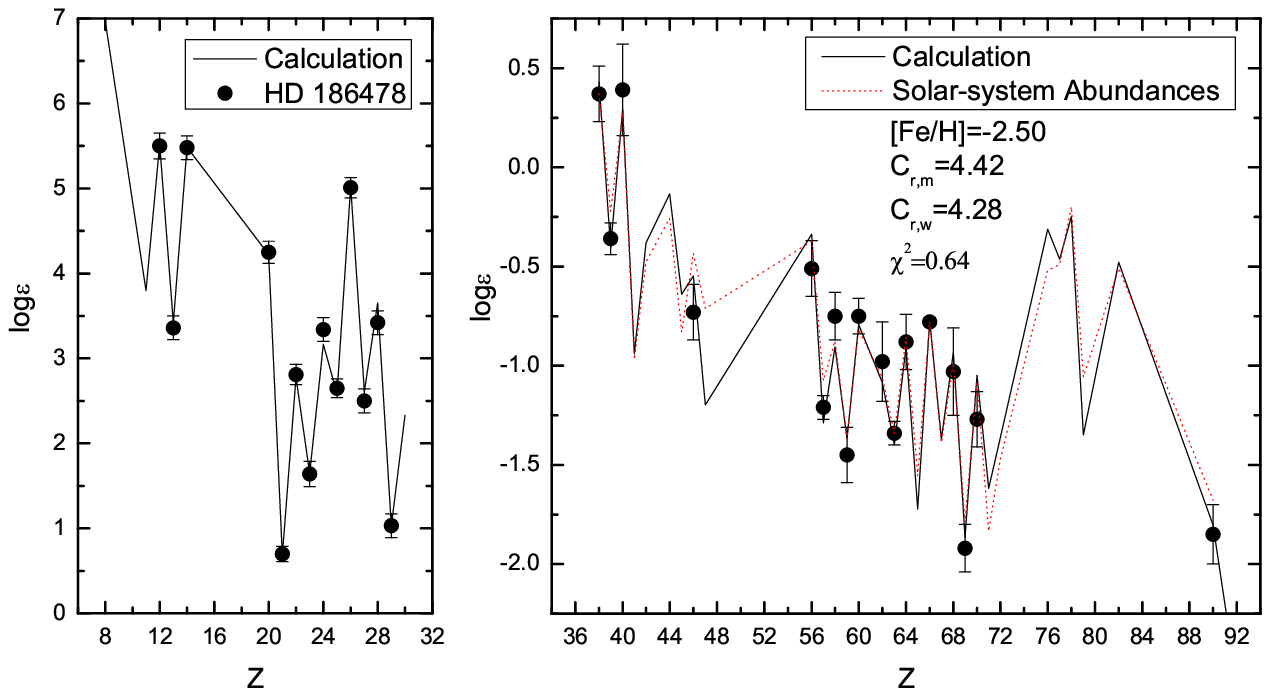}
\caption{Fitted result for the sample star HD 186478. Symbols:
filled circles are observed abundances; the solid line presents the
calculated results and the dotted line is the scaled solar-system
r-process abundances. The left panel is the result for light
elements. The right panel is the result for neutron-capture
elements.}
\end{figure}

\begin{figure}[t]
 \centering
 \includegraphics[width=1\textwidth,height=0.6\textheight]{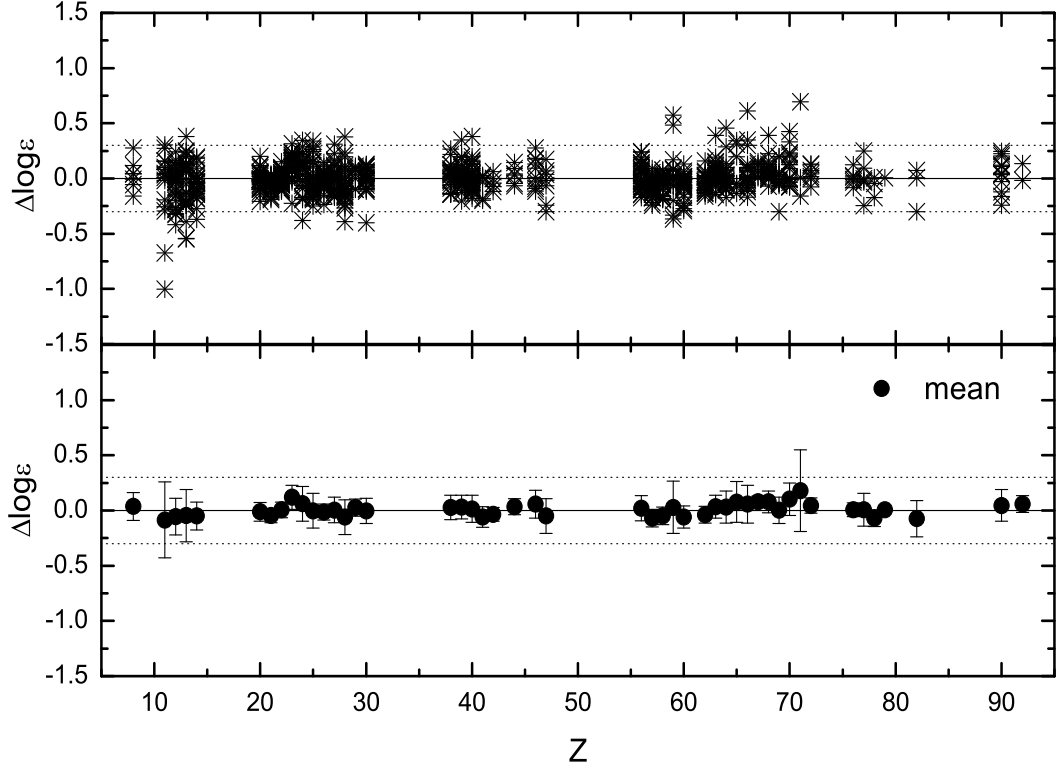}
\caption{Top panel: individual relative offsets
($\Delta\log\varepsilon(X)\equiv \log\varepsilon(X)_{cal}
-\log\varepsilon(X)_{obs}$) for the sample stars (except for BD-18
5550) between the predictions with the observed abundances (stars).
Typical observational uncertainties in $\log\varepsilon$ are
$\sim0.2-0.3$ dex (dotted lines). Bottom panel: The root-mean-square
offsets of these elements in $\log\varepsilon$ (filled circles).}
\end{figure}

\begin{figure}[t]
 \centering
 \includegraphics[width=1\textwidth,height=0.6\textheight]{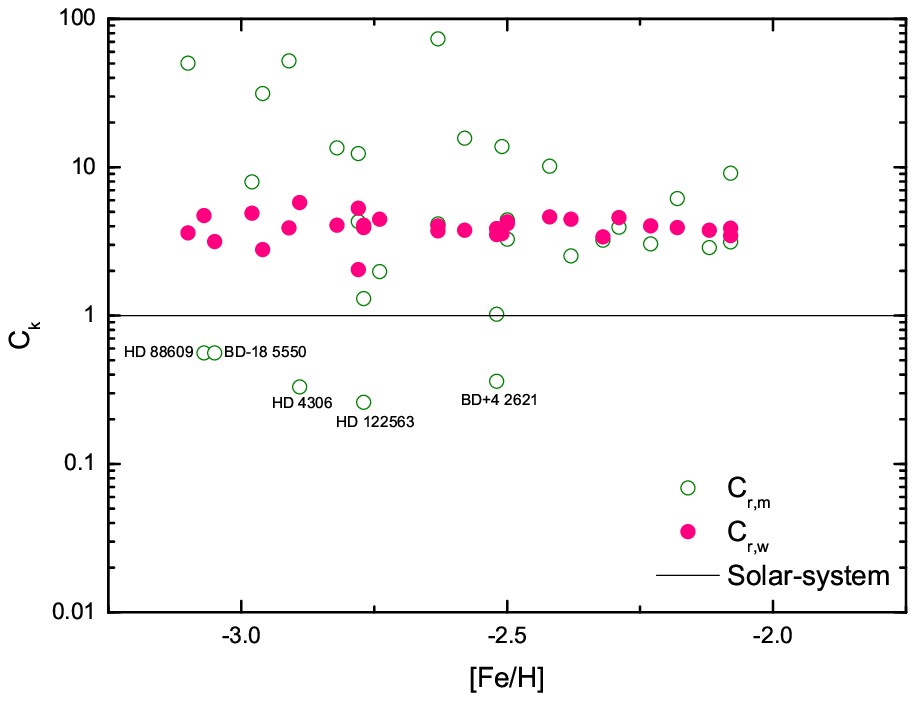}
\caption{The component coefficients as a function of metallicity.
Symbols: open circles and filled circles are the component
coefficients responsible for the main r-process and weak r-process,
respectively; the solid line presents the component coefficients of
the solar system.}
\end{figure}

\begin{figure}[t]
 \centering
 \includegraphics[width=1\textwidth,height=0.6\textheight]{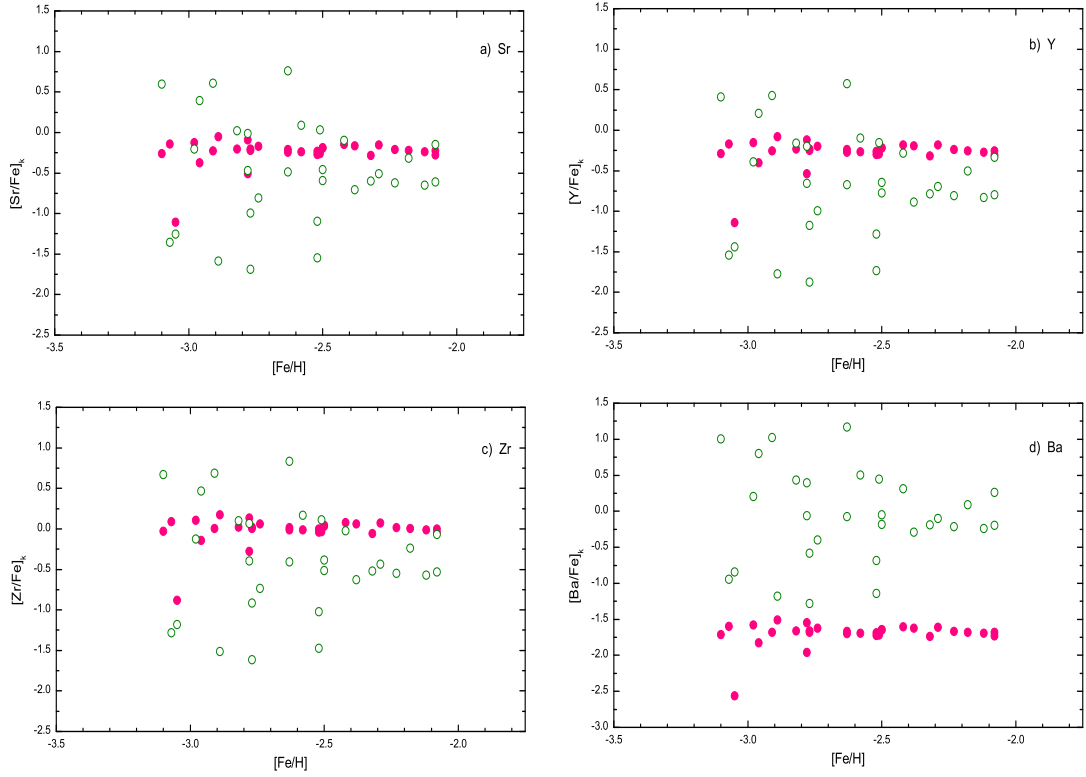}
\caption{Two component ratios of [element/Fe]$_k$ vs. [Fe/H].
Symbols: the open circles and filled circles respectively represent
the main r- and weak r-component ratios calculated in our work.}
\end{figure}

\begin{figure}[t]
 \centering
 \includegraphics[width=1\textwidth,height=0.6\textheight]{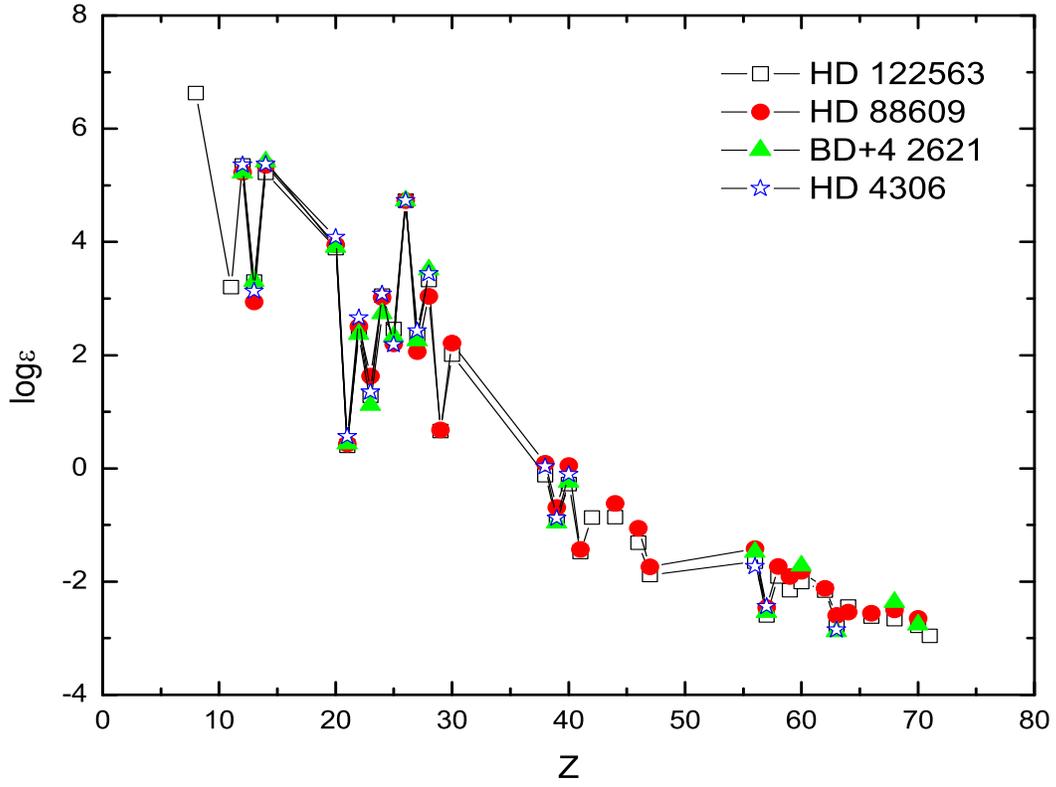}
\caption{The observed abundances of weak r-process stars. Symbols:
open squares, filled circles, filled triangles and open stars are
observed abundances responsible for HD 122563, HD 88609, BD+4 2621
and HD 4306, respectively.}
\end{figure}

\begin{figure}[t]
 \centering
 \includegraphics[width=1\textwidth,height=0.6\textheight]{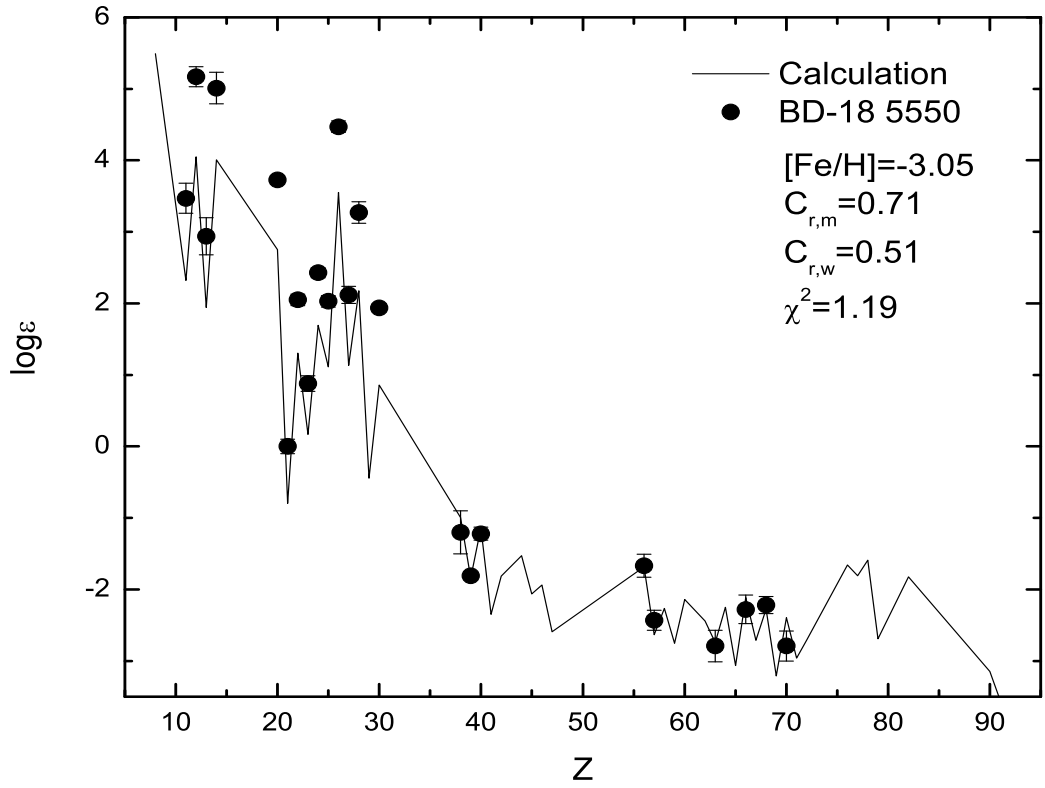}
\caption{The result of BD-18 5550 fitted by two r-process
components. Symbols: solid line and filled circles are the
calculation and observation, respectively.}
\end{figure}

\begin{figure}[t]
 \centering
 \includegraphics[width=1\textwidth,height=0.6\textheight]{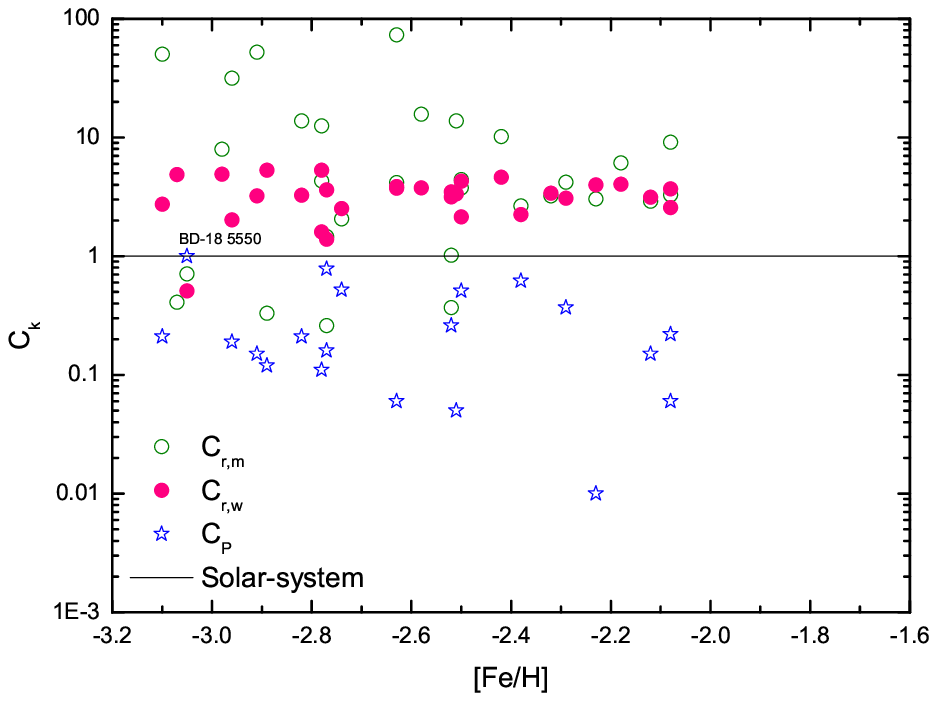}
\caption{The component coefficients as a function of metallicity.
Symbols: open circles, filled circles and open stars are the
component coefficients responsible for the main r-process, weak
r-process and P-component, respectively; the solid line presents the
component coefficients of the solar system.}
\end{figure}

\begin{figure}[t]
 \centering
 \includegraphics[width=1\textwidth,height=0.6\textheight]{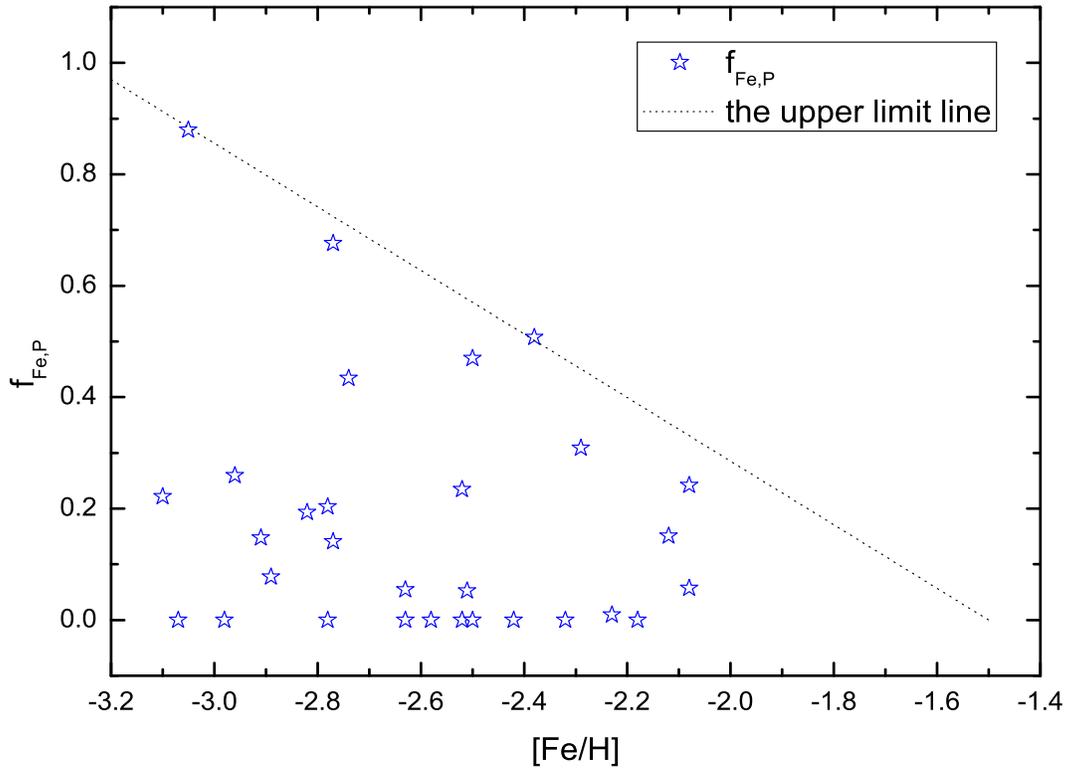}
\caption{The component fractions $f_{i,P}$ for element Fe as
function of [Fe/H]. The dotted line is the upper limit line.}
\end{figure}

\end{document}